\documentclass[english,twoside,11pt]{llncs} 
\usepackage[latin1]{inputenc}
\usepackage{amsmath}
\usepackage{amssymb}
\usepackage{babel}

\newcommand{\bbm}{\begin{bmatrix}}
\newcommand{\ebm}{\end{bmatrix}}
\newtheorem{theo}{Definition}
\newcommand{\tab}{\hspace{0.3cm}}

\def\vec{%
\overrightarrow}

\begin{document} 
\markboth{WLPE 2005}{Likely Invariants Validation with CLP(FD)} 
\pagestyle{myheadings}
\date{} 

\title{Proving or Disproving Likely Invariants \\ with Constraint Reasoning
\thanks{In A. Serebrenik and S. Mu{\~n}oz-Hern{\'a}ndez (editors), 
Proceedings of the 15th Workshop on Logic-based methods in Programming 
Environments, October 2005, Spain. COmputer Research Repository 
(http://www.acm.org/corr/), cs.SE/0508108; whole proceedings: 
cs.PL/0508078.}} 
\author{Tristan Denmat\inst{1} \and Arnaud Gotlieb\inst{2} \and Mireille Ducass\'e\inst{1}}

\institute{IRISA/INSA \and IRISA/INRIA \\ Campus universitaire de Beaulieu 35042 Rennes Cedex, France \email{\{denmat,gotlieb,ducasse\}@irisa.fr}}

\maketitle

\begin{abstract}
A program invariant is a property that holds for every execution of the program.
Recent work suggest to infer {\it likely}-only invariants, via dynamic analysis. A likely
invariant is a property that holds for some executions but is not guaranteed to hold
for all executions. In this paper, we present work in progress addressing the challenging problem of automatically
verifying that likely invariants are actual invariants.  We propose a constraint-based
reasoning approach that is able, unlike other approaches, to
both prove or disprove likely  invariants. In the latter case, our approach provides
counter-examples. We illustrate the approach on a motivating example where
automatically generated likely invariants are verified. 
\end{abstract}

\section{Introduction}

A program invariant is a property that holds over every execution of
the program. Examples of program invariants include loop invariants
presented by Hoare in the weakest precondition calculus \cite{H69} or
pre-post conditions of the design by contracts approach \cite{Mey97}.
Invariants have proved to be crucial in various fields of software
engineering such as specification refinement, software evolution or
software verification.  Unfortunately, writing invariants is a tedious
task and few programmers write program invariants by themselves.

In order to palliate this problem, a trend of research aims at inferring invariants \textit{a posteriori}.
In this case, invariants correspond to the actual behaviors of programs, not to their intended behaviors. 

A common approach is to use static analysis, which infers invariants
from the source code. For example, abstract interpretation-based
analyses generate different kinds of invariants, depending on the
abstract domain used~: intervals \cite{CC77}, polyhedra \cite{CH78} or
octagons \cite{Min04}, to name a few.  These methods generate sound
invariants but the abstractions used to address problems of
termination and complexity may lead to a weak accuracy.

Ernst et al.  introduced Daikon, a tool performing dynamic inference of properties using
actual values computed during program executions \cite{ECGN01}. The
advantage is that the generated properties are in general more precise
than those generated with a static inference. 
The drawback of this method is that the properties may not hold for
particular executions. They are therefore \emph{likely} only invariants.
Proving likely invariants to be correct would make them sound, while
being in general more precise than statically inferred invariants.

In this paper, we present work in progress regarding a
constraint-based approach to verify likely invariants by refutation.
We have restrained the presentation to the validation of likely
invariants generated by Daikon.  Nevertheless, others likely
invariants can be checked with this approach. For example, a user
could test his program against properties that he knows it is supposed
to have.  The idea of the approach is, firstly, to generate a
constraint system, $CS$, modeling an imperative program. To do this,
we use the translation of an imperative program into CLP(FD) presented
by Gotlieb et al. \cite{GBR98}, which has already proved to be useful
in structural testing \cite{GBR00}. This transformation can deal with
a large subset of C/C++ language, including floating point numbers
\cite{BGM05} and a restricted class of pointers \cite{GDB05}. Then, we
transform a likely invariant into a constraint $I$.  Finally, we try
to find a solution of the constraint system $CS \land \lnot I$~: if
the constraint solver finds a solution, then the likely invariant is
spurious. If the solver finds that there is no solution, then the
likely invariant is an invariant.  Unfortunately, the resolution might
not terminate or might take too long. In these cases, nothing can be
concluded. From a declarative point of view, this approach is very
similar to the verification of program based on Horn Logic Denotations
\cite{G99}. The difference is the use of constraint logic to express
the semantics of an imperative language instead of pure Horn logic.
When running the verification, Horn logic leads to a generate-and-test
method, whereas constraint logic leads to a
propagate-generate-and-test method. We expect our approach to be more
efficient because the propagation should reduce the number of test
cases.

The different steps of our approach are detailed on a motivating example.
Three likely invariants are generated by a dynamic inference. By
applying the method presented here, two of them are disproved and the
other is proved.

The contribution of the approach, as illustrated on our example, is to
be able to both prove or disprove some likely invariants.  In the
literature, similar techniques are dedicated to either one or the
other. Jackson and Vaziri use a constraint solving-based approach that
only allows them to disprove likely invariants \cite{JV00}.  Nimmer
and Ernst present an experiment to prove the correctness of likely
invariants using the static checker ESC-Java \cite{BCC05,NE02}.
When ESC-Java fails to prove a likely invariant, it might be due to
the lack of an assertion or precondition rather than to an actual
error.  Because of this point, ESC-Java cannot disprove spurious likely
invariants.

Section 2 briefly describes the work of Ernst et al. on dynamic
inference of likely invariants.  Section 3 presents our motivating
example. The dynamic analysis of Ernst et al. is used to infer
invariants on this program.  Section 4 summarizes the translation of
an imperative program into a constraint system.  Section 5 illustrates
how we suggest to refute or prove a likely invariant using constraint
solving.  Section~6 discusses difficulties encountered with our
approach. Finally, section~7 concludes the paper.

\section{Dynamic inference of invariants}
\label{section:daikon}
This section briefly describes the seminal work of Ernst et al. on
dynamic inference of likely invariants \cite{ECGN01}.

Previous work about the inference of program invariants used static
analyses.  Results of such analyses are sound, which is very important
for program invariants. The counterpart is that the approximations and
complex algorithms required to achieve soundness may lead to a weak
accuracy.

Ernst et al. propose a compromise where the soundness of the results
is not guaranteed in order to gain accuracy. They use dynamic analyses
that compute likely invariants from data collected during executions.
The underlying idea is that, if a property holds over many executions,
then it has good chances to be an invariant.

Daikon is a tool that implements the dynamic inference of likely
invariants in four steps.  Firstly, the program is instrumented to
automatically trace values of variables of interest during execution.
Secondly, a test suite is executed on this new program.  The data
collected during these executions are stored in a database.  Thirdly,
the set of potential likely invariants is generated. Daikon uses a
pool of relationships to automatically generate all potential
invariants between variables that can be compared. Comparability
between variables is discussed in \cite{ECGN00}.  Examples of possible
invariants are equalities with a constant (e.g.  $x = a$), non-linear
relationships among variables (e.g. $z = gcd(x,y)$) or ordering
relationships between variables (e.g. $x > y$). Additional
relationships involving at most three variables are trivial to add.
Finally, the set of possible invariants so-generated is checked
against the execution data stored in the database.  Possible
invariants that are not falsified during this checking are reported to
be likely invariants.

In practice, the complexity of the Daikon algorithm tends to be proportional
to the number of detected invariants. A lot of research is done
around Daikon to improve the efficiency and accuracy of the inference.

\section{Running example}

This section presents an example of dynamic inference of likely invariants
as presented in section~\ref{section:daikon}. Figure \ref{fooC} shows the $foo$ C program. This program takes two
input values~: $n$ and $r$. It returns $r$ if $n$ is negative. Else, it returns $r$ if $n$ is even and $r + 1$ if $n$ is odd.
 
\begin{figure}[t]
\begin{center}  
\begin{tabular}{l}
int foo (int $n$, int $r$)\{ \\
\tab int $s = 0$; \\
\tab while $(n > 0)$ \{   \\
\tab \tab \tab  $n --$;  \\
\tab \tab \tab if $(s == 0)$\{   \\
\tab  \tab \tab \tab \tab $s = 1$;  \\
\tab  \tab \tab \tab \tab $r ++$; \\
\tab  \tab \tab \tab  \} \\
\tab  \tab \tab else \{\\
\tab  \tab \tab \tab \tab $s = 0$; \\
\tab  \tab \tab \tab \tab $r --;$ \\
\tab  \tab \tab \tab \} \\
\tab  \tab \} \\
 \tab return $r$; \\
\tab \} \\
\end{tabular}
\caption{A toy example : the $foo$ program}
\label{fooC}
\end{center}
\end{figure}

We have used Daikon, the tool presented in section
\ref{section:daikon}, to infer likely program invariants of the $foo$
program.  We used an all-branch covering test
suite of 25 test cases. In these test cases, the loop is unfolded from
0 to 454 times. With this test suite, the inference configured in the
default mode resulted in three likely invariants at the exit point of
the program~:
\begin{enumerate}
\item $orig(r) = 0 \implies return = 0$ \label{inv1}
\item $return = 0 \implies orig(r) = 0$ \label{inv2}
\item $return \geq orig(r)$ \label{inv3}
\end{enumerate}

In these likely invariants, $orig(r)$ corresponds to the value of
variable $r$ at the entry point of the program and $return$ is the
value returned by the program. These likely invariants are not
trivial, as they represent a partial specification of a loop. In
particular, likely invariant \ref{inv3} is complicated to infer
statically.  Indeed, it requires to detect that the executed branch of
the conditional alternates at each loop unfolding in such a way that
the value of $r$ cannot become lower than $orig(r)$.  Likely
invariants \ref{inv1} and \ref{inv2} are also difficult to infer as
they can be seen as a disjunction of two properties.  For example,
likely invariant \ref{inv1} is actually $orig(r) \neq 0 \lor return =
0$.

\section{Translation of an imperative program into a constraint system}
\label{section:transformation}

This section describes the first step of our approach to validate
likely invariants, namely translating an imperative program into
constraint logic programming on finite domains (CLP(FD)).  More
details about the transformation can be found in \cite{GBR98}.

CLP(FD) is an extension of logic programming. In CLP(FD) programs,
logical variables are assigned a \emph{domain} and relations between
variables are described with \emph{constraints}. A solution to a
CLP(FD) program is a valuation of every variable in its own domain
such that no constraint is falsified. Solutions are find using two
mechanisms~: \emph{propagation} and \emph{enumeration}. Propagation
uses domain information of each variable to reduce domains of other
variables. When no more propagation can be done, enumeration, also
called labeling, assigns values to variables to find a solution.  Note
that each time a variable is assigned a value, a new propagation phase
takes this new information into account.

The goal of the transformation described in the following is to
generate a CLP(FD) constraint between the input and output variables
of an imperative program. Values for which this constraint is
satisfied are those who correspond to an existing execution of the
program.  More formally, if \emph{In} is the list of input variables
of the program and \emph{Out} the list of output variables, a
constraint $clp\_prog(In,Out)$ is generated. If the pair $(I,O)$ is a
solution of $clp\_prog$ then the execution of the original program on
inputs $I$ returns values $O$.
 
The translation uses the SSA-form as an intermediary form of the
program. The instructions of the intermediary program are transformed
into constraints. In particular, specific operators are designed to
deal with control structures.

\subsection{The SSA-form}
\label{subsection:ssa}
The SSA-form is an intermediate representation of imperative programs
which prepares the translation into CLP(FD). It has originally been
presented by Cytron et al.  to optimize compilers \cite{CFR91}.  The
SSA form is a semantically equivalent version of a program where each
variable has a unique definition and every use of this variable is
reached by the definition.

The SSA-form is relevant here because logical variables in CLP(FD)
programs can be assigned only once whereas, in imperative programs not
in SSA-form, variables can be assigned many times.

Every program can be transformed into SSA by renaming the uses and
definitions of the variables. For example $i = i+1; j = j*i$ is
transformed into $i_2 = i_1 + 1; j_2 = j_1 * i_2$.  At the junction
nodes of the control structures, SSA introduces special assignments
,called $\phi$-functions, to merge several definitions of the same
variable~: $\vec{v_2} = \phi(\vec{v_0},\vec{v_1})$ assigns the values
of $\vec{v_0}$ in $\vec{v_2}$ if the flow comes from the first branch
of the decision, $\vec{v_1}$ otherwise. In the case of conditional
structures, $\vec{v_0}$ and $\vec{v_1}$ are respectively the vectors
of defined variables in the \emph{then} and \emph{else} branches.
$\vec{v_2}$ is the vector of these variables out of the conditional
structure.  Depending on the validity of the condition, $\vec{v_2} =
\vec{v_0}$ or $\vec{v_2} = \vec{v_1}$.

\subsection{Instructions as CLP(FD) constraints}
\label{subsection:translating}

The instructions of the original program are transformed into
constraints between logical variables.  Type declarations are
translated into domain constraints. For example, the declaration of a
signed integer $x$ is translated into~: $X \in -2^{31}..2^{31}-1$
where $X$ is a logical FD\_variable.

Assignments and decisions are translated into arithmetical
constraints. For example, assignment $x=x+1$ is converted into the SSA
form $x_2 = x_1 + 1$ and further translated into $X_2 = X_1 + 1$ where
$X_1,X_2$ are logical FD\_variables.

The main difficulty is to transform control structures into
constraints. As described in the following, two specific operators are
used.

\subsubsection{Conditional statements} 
The conditional statement is treated with a specific combinator {\tt
  ite/6}.  Arguments of {\tt ite/6} are the variables that appear in
the $\phi$-functions and the constraints generated from the different
parts of the original conditional statement.  Note that other
combinators may be nested into the arguments of {\tt ite/6}.  The SSA
{\bf if\_else} statement~:
\begin{displaymath}
{\bf if} ({\it exp})~{\bf \{ }{\it stmt} {\bf \}}~{\bf else}~{\bf \{ } {\it stmt} {\bf \}}~\vec{v_2} = \phi(\vec{v_0}, \vec{v_1})
\end{displaymath}
is translated into ${\tt ite}(C_{Cond}, \vec{v_0}, \vec{v_1},
\vec{v_2},C_{Then}, C_{Else}) $ where $C_{Cond}$ is a constraint
generated by the analysis of {\it exp} and $C_{Then}$ (resp.
$C_{Else}$) is a set of constraints generated by the analysis of the
\emph{then} branch (resp.  \emph{else} branch).

The combinator {\tt ite/6} is defined as~:
\begin{theo}{\tt ite/6} 
\begin{tabbing}
{\tt ite}($C_{Cond},$\=$ \vec{v_0}, \vec{v_1}, \vec{v_2}, C_{Then}, C_{Else}):- $\\
\> $C_{Cond} \longrightarrow C_{Then} \wedge \vec{v_2} = \vec{v_0}$,\\
\>$\neg C_{Cond} \longrightarrow C_{Else} \wedge \vec{v_2} = \vec{v_1}$,\\
\>$\neg(C_{Cond} \wedge C_{Then} \wedge \vec{v_2} = \vec{v_0}) \longrightarrow \neg C_{Cond} \wedge C_{Else} \wedge \vec{v_2} = \vec{v_1}$,\\
\>$\neg(\neg C_{Cond} \wedge C_{Else} \wedge \vec{v_2} = \vec{v_1}) \longrightarrow C_{Cond} \wedge C_{Then} \wedge \vec{v_2} = \vec{v_0}$,\\
\>$(C_{Cond} \wedge C_{Then} \wedge \vec{v_2} = \vec{v_0}) \veebar  (\neg C_{Cond} \wedge C_{Else} \wedge \vec{v_2} = \vec{v_1}).$
\end{tabbing}
\end{theo}

This definition uses \emph{guarded-constraints}. A guarded-constraint
$head~\longrightarrow~tail$ rewrites into $tail$ if the constraint
$head$ is entailed by the constraint store.  The first two
guarded-constraints straightforwardly result from the operational
semantics of the {\bf if\_else} statement whereas the third and the
fourth correspond to a backward reasoning. In this case, values of
$\vec{v_2}$ are used to deduce information concerning the flow.  The
last constraint contains the constructive disjunction operator
$\veebar$. This operator removes from the domains of the variables the
values that are removed whatever the executed part of the disjunction
is. For example, if the constraint {\tt ite}$(..., \bbm X_0 \ebm, \bbm
X_1 \ebm, \bbm X_2 \ebm, X_0 = 1, X_1 = 3)$ stands, the constructive
disjunction operator deduces that $X_2 \in
\{1,3\}$.

\subsubsection{Iterative statements} 

The SSA {\bf while} statement 
\begin{displaymath}
\vec{v_2} = \phi(\vec{v_0},\vec{v_1})~{\bf while}({\it exp})~{\bf \{ } {\it stmt} {\bf \} }
\end{displaymath} 
is treated with the recursive specific combinator  {\tt w}$(C_{Cond},\vec{v_0},\vec{v_1},\vec{v_2}, C_{Body})$ where $C_{Cond}$ is a constraint generated by the analysis of {\it
  exp} and $C_{Body}$ is a set of constraints
generated by the analysis of \emph{stmt}.

\newpage
\begin{theo}{\tt w/5} 
\begin{tabbing}
{\tt w}$(C_{Cond}$\=$,\vec{v_0},\vec{v_1},\vec{v_2},C_{Body}):-$\\ 
\>  $C_{Cond} \longrightarrow ( C_{Body} \wedge {\tt w}(C'_{Cond},\vec{v_1},\vec{v_3},\vec{v_2}, C'_{Body}))$,\\
\>  $\neg C_{Cond} \longrightarrow \vec{v_2} = \vec{v_0}$,\\
\>  $\neg (C_{Cond} \wedge C_{Body}) \longrightarrow (\neg C_{Cond} \wedge \vec{v_2} = \vec{v_0})$,\\
\>  $\neg ( \neg C_{Cond} \wedge \vec{v_0} = \vec{v_2}) \longrightarrow ( $\=$C_{Cond} \wedge C_{Body} 
                \wedge $\\\>\>${\tt w}(C'_{Cond},\vec{v_1},\vec{v_3},\vec{v_2}, C'_{Body})).$ 
\end{tabbing}
\end{theo}

Note that combinator {\tt w/5} is dynamic~: new variables and new
constraints are generated during its evaluation.  In particular, the
vector $\vec{v_3}$ is a vector of fresh variables. The first and the
last guarded constraints both make a recursive call to $w$. The
parameters of this new $w$ are not $C_{Cond}$ and $C_{Body}$ but new
constraints $C'_{Cond}$ and $C'_{Body}$ where some variables have been
substituted by variables of $\vec{v_1}$ and $\vec{v_3}$ to model the
fact that the loop has already been entered once.

The first two guarded--constraints are deduced from the operational
semantics of the {\bf while} statement.  The third constraint tells
that, if the constraints extracted from the body are proved to be
contradictory with the current constraint system then the loop cannot
be entered. The last constraint models the fact that, if any variable
possesses distinct values before and after the execution of the {\bf
  while} statement, then the loop must be entered at least once.

\subsection{Translation of the $foo$ program into constraints}
\label{subsection:exempleJouet}

This section presents the translation of the $foo$ program of
Figure~\ref{fooC} into a constraint system. By applying the
translation described above, the constraint system presented in Figure
\ref{progCLP} is generated.

\begin{figure}[t]
\begin{center}
  \begin{tabular}{|l|l|}
\hline
& \\
    \begin{tabular}{l}
int foo (int $n$, int $r$)\{ \\
\tab int $s = 0$; \\
\tab while $(n > 0)$ \{   \\
\tab \tab \tab  $n --$;  \\
\tab \tab \tab if $(s == 0)$\{   \\
\tab  \tab \tab \tab \tab $s = 1$;  \\
\tab  \tab \tab \tab \tab $r ++$; \\
\tab  \tab \tab \tab  \} \\
\tab  \tab \tab else \{\\
\tab  \tab \tab \tab \tab $s = 0$; \\
\tab  \tab \tab \tab \tab $r --;$ \\
\tab  \tab \tab \tab \} \\
\tab  \tab \} \\
 \tab return $r$; \\
\tab \} \\
\end{tabular}
&
\begin{tabular}{p{1 cm} l}
\multicolumn{2}{l}{foo([$N_0,R_0$],[$RET$]):-} \\
& $S_0 = 0$, \\
& w($n > 0$,$\overrightarrow{V_{old}}$,$\overrightarrow{V_{new}}$,$\overrightarrow{V_{final}}$, \\
& \tab [$n = n - 1$, \\
& \tab ite($s = 0$, $\overrightarrow{V_{then}}$,$\overrightarrow{V_{else}}$,$\overrightarrow{V_{f\_ite}}$, \\
& \tab \tab [$s = 1, r = r + 1$], \\
& \tab \tab[$s = 0, r = r - 1$])]), \\
& $RET = R_{final}$. \\
\end{tabular}
\\
& \\
\hline
\end{tabular}
\caption{Translation of the $foo$ program into a constraint system}
\label{progCLP}
\end{center}
\end{figure}

For the sake of clarity, we omit the translation into SSA-form. That
is why the constraint system presented on Figure \ref{progCLP} does
not explicitly show all the SSA-names. In fact, the variable names
that are in the parameters of the $w$ and $ite$ operators must be
considered only as syntactical names. Depending on the cases, these
names are replaced by logical variables that are in the vectors
$\overrightarrow{V_{old}}$,$\overrightarrow{V_{new}}$,$\overrightarrow{V_{final}}$,$\overrightarrow{V_{then}}$,$\overrightarrow{V_{else}}$
or $\overrightarrow{V_{f\_ite}}$. Constraints that correspond to the
type declarations of variables are also omitted.

As the transformation faithfully models the operational semantics of C
programs, the constraint system can be executed just like the original
C program. For example, if we instantiate $N_0$ to $5$ and $R_0$ to
$3$, constraint propagation leads to the instantiation of $RET$ to
$4$, which is the result of the original program on the same entries.

\section{Validation of likely invariants} 

In this section, we informally introduce a method to prove or disprove
likely invariants. Section~\ref{validation:constraints:section}
explains how we transform the problem of invariant validation into a
constraint satisfaction problem and
Section~\ref{subsection:examplevalidation} illustrates the behavior of
constraint solvers for the running example.

\subsection{A constraint solving problem}
\label{validation:constraints:section}

Section~\ref{section:transformation} presented a model of an
imperative program as a constraint system. This constraint system,
denoted by $CS$, is a relation between the input variables and the
output variables. If $(X,Y)$ is a solution of $CS$, $X$ and $Y$ being
respectively input and output values, then there exists a finite
execution of the original program starting with input $X$ and
returning $Y$.

A likely invariant, denoted by $I$, can be seen as one more
constraint. This new constraint should be implied by $CS$ if $I$
really is an invariant. We want to prove
\begin{equation}
CS \vDash  I
\nonumber
\end{equation}

Such a proof can be established by refutation using constraint solving~: 

\begin{equation}
CS \vDash I \Leftrightarrow Sol(CS \land \lnot I) = \emptyset
\nonumber
\end{equation}

In this equation, $Sol(CS \land \lnot I)$ denotes the set of solutions of the constraint system $CS \land \lnot I$.  

When solving the refutation request $CS \land \lnot I$, there are three cases~:
\begin{enumerate}
\item there exists a solution $(X,Y)$, which means that the execution
starting from $X$ and terminating in $Y$ does not verify the likely invariant
$I$. Thus, $I$ is spurious and $(X,Y)$ is a counter-example.
\item there is no solution to this problem.  It means that $I$ really is an invariant.
\item the user runs out of patience. It can be due either to a too
  long computation or a non-terminating computation. Nothing can be concluded.
\end{enumerate}

As already mentioned in the introduction, the method presented by
Nimmer and Ernst~\cite{NE02} can prove that a program verifies a
likely invariant.  However, if no proof can be established, it might
be due to the fact that there is not enough axioms. For example, loop
invariants must be provided by users in order to soundly prove
properties~\cite{BCC05}.
On the contrary, in the work of Jackson and Vaziri \cite{JV00}, it is
possible to find a counter-example that does not verify the
property. However, if none can be found, it can be due either to
the fact that the likely invariant is indeed an invariant or to the
inaccuracy of the under-approximation. For example, as the number
of loop unfoldings is bounded by a value $k$, there might exists a
counter-example that unfolds $k + 1$ times a loop.

In other words, at the question \textit{does the program verify the
property~?}, Nimmer and Ernst answer ``yes'' or
``maybe'', Jackson and Vaziri answer ``no'' or ``maybe'' and our
method answers ``yes'', ``no'' or ``maybe''. 

\subsection{Validation of the invariants of the running example}
\label{subsection:examplevalidation}

In this section, we illustrate our approach on the running example.
The first likely invariant inferred by Daikon for the $foo$ program is
\begin{equation}
orig(r) = 0 \implies return = 0.
\nonumber
\end{equation}
As explained in the previous paragraph, the first step of the
validation consists in adding the negation of the likely invariant to
the program. The request sent to the solver is therefore
\begin{equation}
:- foo([N_0,R_0],RET), R_0 = 0, RET~\setminus=0.
\end{equation}
After propagation the solver answers :
\begin{equation}
N_0 \in [inf,sup], RET \in [inf,-1] \cup [1,sup], R_0 = 0
\end{equation}
The propagation alone did not allow the solver to find inconsistencies in
the constraint system. Nothing can be deduced concerning the
invariant unless concrete values for $N_0$ and $RET$ are found. An
enumeration step on variables $N_0$ and $RET$ must be done. Note that
variables need to have a domain for labeling. As the logical variables
correspond to integers in the original imperative program, 
their bounds are $MIN\_INT$ and $MAX\_INT$. The request is now~:
\begin{multline}
:- domain([N_0,RET],MIN\_INT,MAX\_INT), foo([N_0,R_0],RET), \\R_0 = 0, RET ~\setminus =0, labeling([N_0,RET]).
\end{multline}
After propagation and enumeration, the solver finds a solution 
\begin{equation}
N_0 = 1, R_0 = 0, RET = 1. 
\end{equation}
It means that the execution of the original program with input $n=1,
r=0$ returns $ret = 1$. This execution is a counter-example of the
likely invariant $orig(r) = 0 \implies return = 0$. It is therefore
disproved.\\

The second likely invariant inferred by Daikon for the $foo$ program is
\begin{equation}
return = 0 \implies orig(r) = 0.
\nonumber
\end{equation}
In the same way as above, a counter-example is found~: 
\begin{equation}
n=1, r= -1, return = 0.
\nonumber
\end{equation}
The second likely invariant is therefore also disproved.\\

The third likely invariant inferred by Daikon for the foo program is
\begin{equation}
return \geq orig(r).
\nonumber
\end{equation}
Repeating the operations previously detailed, the following
request is sent to the constraint solver~:
\begin{equation}
:- foo([N_0,R_0],RET), 
\nonumber
\end{equation}
\begin{equation}
R_0 > RET.
\label{hypo:3e:invariant}
\end{equation}
This time, without any enumeration, the constraint solver answers ``no'',
meaning that there is no solution to this problem. The third likely
invariant is therefore proved to be an invariant.

The behavior of the {\tt w} operator on the latter refutation is as
follows.

Initially, the {\tt w} operator is instantiated to 
\begin{equation}
{\tt w}(N_0 > 0, [R_0,N_0,S_0],[R_1,N_1,S_1], [RET,N_2,S_2],C_{Body})
\nonumber
\end{equation}
We have not expanded the constraint system of the body for readability
reasons.  The fourth guarded constraint of the {\tt w} operator
instantiated for the {\tt foo} program is logically equivalent to what
follows.

\begin{multline} 
N_0 > 0 \lor (R_0 \neq RET) \longrightarrow ( N_0> 0 \wedge C_{Body}
\wedge  \\{\tt w}(N_1 > 0, [R_1,N_1,S_1], [R_3,N_3,S_3],[RET,N_2,S_2], C'_{Body}))).  
\nonumber
\end{multline}

As $R_0 > RET$ (constraint \ref{hypo:3e:invariant}), it is impossible
for $R_0$ to be equal to $RET$. The guard of the previous constraint
is entailed.
The loop must therefore be entered and constraints of $C_{Body}$ are set up
\begin{equation}
N_1 = N_0 - 1.
\label{decrement:N:contrainte}
\end{equation}
As $S_0 = 0$ (first constraint of
the foo program), the {\tt ite} operator set up constraints
corresponding to the \emph{then} branch
\begin{equation}
S_1 = 1
\end{equation}
\begin{equation}
R_1 = R_0 + 1
\label{increment:boucle:contrainte}
\end{equation}
Due to constraints~\ref{hypo:3e:invariant}
and~\ref{increment:boucle:contrainte} the following property is true
\begin{equation}
R_1 > R_0 > RET, 
\end{equation}
therefore, it is impossible to have $R_1 = RET$. Consequently, the loop is
unfold again.  Values $[R_3,N_3,S_3]$ are constrained by clones of
constraints~\ref{decrement:N:contrainte}
and~\ref{increment:boucle:contrainte}. The same reasoning applies
until propagation deduces that $n$ cannot be greater than 0. At the
beginning, $n$ is in the interval $[MIN\_INT,MAX\_INT]$ so after
$MAX\_INT$ iterations $n$ is in the interval $[MIN\_INT,0]$ because of
constraint~\ref{decrement:N:contrainte} and all its clones. Thus,
\begin{equation}
N_{MAX\_INT} \leq 0
\label{fin:boucle:contrainte}
\end{equation}
At this point, $R_{MAX\_INT} > RET$.  The second guarded-constraint of
the {\tt w} operator instantiated for the {\tt foo} program is :
\begin{equation}
\lnot N_k>0 \longrightarrow [R_k,N_k,S_k] = [RET,N_2,S_2]
\nonumber
\end{equation}
When $k = MAX\_INT$, the guard is entailed because of
constraint~\ref{fin:boucle:contrainte}. Consequently, the constraint
\begin{equation}
RET = R_{MAX\_INT} 
\end{equation}
is set up. It makes the constraint store unsatisfiable, and this is
detected by the constraint solver.  As a consequence, the third
invariant is proved to be true.

\section{Discussion}

The previous section presented three examples of validation of likely
invariants by constraint solving.  Two likely invariants were disproved by the exhibition of a counter-example.
The last one was proved to be an invariant.  

A point that we have not developed yet is the case where the
resolution does not terminate or is too long.  There are two main
reasons why these cases can happen. The first reason is due to the
loops. Indeed, as the model we use describes the operational semantics
of a program, if the original program does not terminate, then the
resolution will not terminate.

The second reason is a problem of propagation in the constraint
system.  As presented in section \ref{section:transformation}, the
operators $ite$ and $w$ are defined via guarded-constraints.
Consequently, if the entailment of none of the guards can be deduced
from the current store of constraints, then the resolution of the
constraint system suspends.  The problem is that our system is very
specific and usual methods of entailment-checking are inefficient in
this context~: domains of variables are very large, constraint store
is dynamic and constraints used can be non-linear.

The consequence of this lack of propagation is that, in bad cases,
almost all the possible values of input variables will have to be
enumerated to prove or disprove likely invariants. In such a case, our
approach becomes a generate-and-test method, which is intractable when
the domains of input variables are large.  Future work will consist in
improving the propagation inside our specific constraint system.
 
\section{Conclusion}
In this paper, we have presented an approach to verify the correctness
of likely invariants using constraint solving. We have illustrated its
principles on a toy example.

The originality of this method is that some likely invariants are
disproved and others are proved. This differs from other methods that
are dedicated to only one of these capabilities.  Methods using
under-approximations can only disprove likely invariants whereas
methods using over-approximation can only prove likely invariants. We
are not using any approximation, it allows us to prove and disprove
but prevents us to guarantee termination and good performances.
Consequently, the key point of our approach is to have a good
propagation inside the constraint system in order to reduce as much as
possible the number of cases where we cannot conclude.

\paragraph{Acknowledgments} We thank the anonymous referees for their helpful comments.

\end{document}